# Macro- and microscopic properties of strontium doped indium oxide

Y.M. Nikolaenko[1], Y.E. Kuzovlev[1], Y.V. Medvedev[1], N.I. Mezin[1],
C. Fasel[2], A. Gurlo[2], L. Schlicker[2], T.J.M. Bayer[2], Y.A. Genenko[2]

[1] *Donetsk Institute for Physics & Technology, National Academy of Sciences of Ukraine, 83114 Donetsk, Ukraine*

[2] *Institute of Materials Science, Darmstadt University of Technology, 64287 Darmstadt, Germany*

**Abstract**

Solid state synthesis and physical mechanisms of electrical conductivity variation in polycrystalline, strontium doped indium oxide $In_2O_3:(SrO)_x$ were investigated for materials with different doping levels at different temperatures (T=20-300$^0$C) and ambient atmosphere content including humidity and low pressure. Gas sensing ability of these compounds as well as the sample resistance appeared to increase by 4 and 8 orders of the magnitude, respectively, with the doping level increase from zero up to $x$=10%. The conductance variation due to doping is explained by two mechanisms: acceptor-like electrical activity of Sr as a point defect and appearance of an additional phase of $SrIn_2O_4$. An unusual property of high level ($x$=10%) doped samples is a possibility of extraordinarily large and fast oxygen exchange with ambient atmosphere at not very high temperatures (100-200$^0$C). This peculiarity is explained by friable structure of crystallite surface. Friable structure provides relatively fast transition of samples from high to low resistive state at the expense of high conductance of the near surface layer of the grains. Microscopic study of the electro-diffusion process at the surface of oxygen deficient samples allowed estimation of the diffusion coefficient of oxygen vacancies in the friable surface layer at room temperature as $3\times10^{-13}$ cm$^2$/s, which is by one order of the magnitude smaller than that known for amorphous indium oxide films.

## 1. Introduction

In spite of wide applications of indium oxide based materials many physical aspects concerning their conductivity mechanism are still disputed. Recent advance in understanding of band structure details was reviewed in Ref. [1]. According to conventional interpretation of optical data the fundamental energy gap of wide band semiconductor $In_2O_3$ was assumed to be a value >3.6 eV. The smaller gap of about 2,7eV was attributed to indirect electron transitions, but the careful studies in Ref. [2] suggested that the indirect gap is not a property of the bulk $In_2O_3$. Furthermore, on the basis of theoretical calculations and hard x-ray photoemission spectra the authors of Ref. [1] concluded a new scheme of band structure. According to this the minimal direct gap between valence and conduction bands corresponds to 2.9 eV. Peculiarity of existing optical data was explained by the fact that electron transition is forbidden by an optical selection rule at the top of this valence band. The top of the second valence band is located about 0.8 eV below, so that the visible energy gap up to the conduction band bottom corresponds to energy 3.75 eV.

In a later paper [3] the existence of the direct gap of 2.7±0.1eV was derived from analysis of photoemission data for macroscopic single crystals of pure indium oxide. Thereby, the corrections in band structure become feasible for explanation of various experimental data, but they still do not resolve another problem - the absence of the dielectric state in samples of indium oxide at room temperature, expected from the wide band gap. This problem manifests itself in investigations of materials of different kinds and quality, including polycrystalline samples [4], single crystals [3], and high quality films [5,6]. Hence, attention should be focused on the electrical activity of surface states and crystal defects. First-principle theoretical calculations of characteristics of intrinsic point defects in $In_2O_3$ are presented in recent papers [7,8]. Many experimental works are focused on the decreasing density of crystals defects in the samples. Different substrate conditions and special treatments, like an oxygen plasma surface treatment, were shown to have crucial effect on the crystal and film growth and their conducting properties [3,5,9]. Some works were focused on the structure of surface [10] and on the difference between properties of surface and sample bulk [6,11,12].

Application aspects and investigations of general properties of bulk samples and different films based on indium oxide are widely presented in the literature. The indium tin oxide (ITO) films must be mentioned first. This compound has wide applications in optoelectronics as highly electrically conducting and transparent for light material [13]. Many current works concerning ITO are focused on improvement of film properties. The authors of Refs. [14,15] investigated doping peculiarities and revealed that major part (2/3) of Sn atoms is not electrically active at the doping level of indium oxide 6%. The relation between electrical activity and segregation of dopant atoms at the thick (0.25 nm) surface layer of nanocrystals was revealed by authors of Ref. [16]. Particularly, a new potential ITO application must be mentioned [17], where authors revealed an effect of third harmonic generation in terahertz frequency band.

Furthermore, amorphous $In_2O_3$ films were intensively investigated during nineties. Practical interest of these studies is explained by application of indium oxide based materials as inorganic sensor for gas analyzers and humidity meters. For improvement of sensor characteristics the doping of indium oxide by different dielectric oxides, such as $Al_2O_3$, $SiO_2$, $SeO_2$, was undertaken [18-20]. Single crystal samples were performed as nanowhiskers with a microscopic length [21,22], nanowires [23,24], nanocones and stripes with very small thickness and width of several nanometers [25]. Investigations of microscopic size plates are presented in a recent paper [26].

All the mentioned compounds demonstrate *n*-type conductivity at room temperature. This is in agreement with the theoretical explanation [8] which suggests that only *n*-type conductivity is possible in $In_2O_3$ at normal conditions. A special condition for realization of *n-p* transition in ITO film was found by authors of Ref. [27] at high temperature $T>950^0C$. Minimum concentration of free electrons in polycrystalline samples prepared by using very pure components is not less than $10^{17}$ cm$^{-3}$ [4]. Doping of indium oxide by Sn can increase the density of free electrons up to $10^{21}$ cm$^{-3}$. Consistent increase of free electron density causes increase of the optical energy gap between the Fermi level and the second valence band top (Burstein–Moss shift) in a wide range up to 4 eV [28].

Another peculiarity of indium oxide based materials is relatively easy variation of the oxygen content in samples by means of the thermal treatment procedure (TTP) at not too high temperatures (250-300$^0$C) in ambient gas atmosphere with different partial pressures of oxygen [24,26]. Oxygen deficiency leads to an increase of the free electron density. This agrees with the well known property of oxygen vacancies as shallow donor centers, supported by theoretical understanding [8]. In this respect the investigations of the metal-insulator transition in amorphous films [29] and small size single crystal samples under consistent changing of the oxygen content [26] should be mentioned.

Doping of polycrystalline indium oxide by Sr in contrast to Sn decreases conductivity and allows reaching a highly resistive state [30]. Consistent decrease of oxygen content by TTP in different ambient gas atmospheres causes transition of the sample into a low resistive state. The highly resistive state formation during oxidation procedure proceeds by two mechanisms: decrease of the free carrier density with increase of the oxygen content and decrease of transparency of potential barriers at the crystallite boundaries [30]. Variation of the potential barrier transparency makes the dependence of resistance on oxygen content significantly stronger. This property allows consideration of Sr doped indium oxide as a prospective material for gas sensing applications.

The last conclusion is supported in this work on the examples of thick films and polycrystalline samples of In$_2$O$_3$:(SrO)$_x$. The studies give an insight in the conductivity mechanism in polycrystalline In$_2$O$_3$:(SrO)$_x$. We reveal the conditions for appearance of strong inhomogeneous conductivity in the bulk samples after partial extraction of oxygen from the polycrystalline sample as well as nano-size structural modifications of grain surfaces, which cause irreversible variation of conductivity. We find a strong switching effect of resistance in an asymmetric two-electrode system, consisting of a small area contact and a large area contact, which is caused by migration of oxygen vacancies in the vicinity of the small area contact under the action of electric field. Investigation of temporal characteristics of the switching effect and numerical modeling of the process allow estimation of the mobility of oxygen vacancies in the friable surface layer of grains.

The paper is organized as follows. In section 2 experimental details of materials preparation and measurement technique involved are disclosed. General structural and electric characterization of differently doped materials is presented in section 3. Specific studies of electric properties during oxidation-reduction of thick films in different atmospheres are described in section 4. Section 5 is devoted to experimental investigation of surface conduction and local resistivity on the surface of bulk samples. The results of section 5 are interpreted by means of a nonlinear drift-diffusion model for oxygen vacancies which is developed and solved in section 6. The overall discussion of the properties of the Sr-doped indium oxide revealed in the current study is given in section 7 and conclusions are summarized in section 8.

### 2. Experimental details

Polycrystalline samples of the strontium doped indium oxide were prepared by a solid state processing described in details in Ref. [30]. Its main advantages are the relative simplicity and cost-efficiency. Differently doped samples were obtained by mixing fine grained powder of pure indium oxide with $SrCO_3$ powder in distilled water. The mixture was further dried and compacted to disks of a 10 mm diameter and 3 mm thickness and subjected to thermal treatment at temperature $1200^0C$ during 10 hours. Weight content of Sr derived from SrO quantity varied in the series of studied samples as $x$=0.25, 0,5, 1, 2, 4, 6, 8 и 10 %.

A phase content of the compounds was investigated by X-ray spectroscopy (XRD). XRD powder diffraction data were collected in flat sample transmission geometry on a STOE STADI P diffractometer with Mo Kα1 radiation and position sensitive detector with a 6° aperture.

For investigation of a microstructure of the samples the scanning electron microscopy (SEM) was used (JSM 6490 LV, by JEOL, Japan). The equipment of microscope included also two onboard spectrometers for energy dispersive analysis of X-ray radiation (EDX) from the sample (model INCA Energy-350 and model INCA Wave-500). This provided service for chemical element microanalyses in the surface layer of the sample with the depth 1-5 μm and, particularly, allowed identification of Sr distribution over the surface.

Electric conductance of the samples was measured by galvanic method using 4-point technique. For the electric contact preparation a silver paint (Electrolube, Germany) was used. The problem of such method of electrode preparation consists in strong influence of silver paint solder on the sample resistance. To avoid this artefact the products of solder decomposition had to be removed from the sample with the help of thermal annealing.

For direct control of oxygen content variation in $In_2O_3:(SrO)_x$ material during thermal treatment the thermogravimetric investigations were performed by a simultaneous thermogravimetric analyzer STA 449C Jupiter® (Netzsch Gerätebau GmbH, Selb/Bavaria). Preliminary the polycrystalline sample was annealed at $300^0C$ for 12 hours in argon atmosphere and, after that, for 12 hours in oxygen atmosphere. The samples were heated up to $300^0C$ with a heating rate of 10 K/min and a holding time of 30 hours at elevated temperature under continuous argon flow of about 25 ml/min. The mass change could be determined by NETZSCH Proteus® software for thermal analysis in relation to previously conducted correction measurements.

To test sensitivity of the material to different atmospheres in view of possible sensor applications the electrical measurements were performed on thick film samples. For thick film preparation the polycrystalline bulk sample (x=10%) was first milled and then the obtained powder was deposited onto a sapphire substrate with four platinum stripe contacts. Deposition was performed using a liquid spirit solution. After deposition the powder sample of 50 μm thickness was annealed at temperature T=500 C in air atmosphere for 3 hrs. Sensitivity of the thick film resistance to the oxygen content of gas atmosphere was determined using homemade equipment. It provided measurements of the sample resistance at temperature 300°C in gas flow atmosphere (100 $cm^3$/s) an allowed fast changing the gas content.

For measurements of local resistivity at the surface of bulk samples an electrical circuit consisting of a small top contact (a pressed metal tip of stainless steel, tungsten or gold) and a wide bottom plate contact was used. The electrical circuit provided measurements in the constant voltage regime and a possibility of step-like switching of voltage polarity.

## 3. Characterization of the polycrystalline samples with different doping level $In_2O_3:(SrO)_x$

Relatively minor range of changes in conductivity of polycrystalline samples of pure indium oxide with variation of oxygen pressure in ambient gas atmosphere is related, as indicated above, to the concentration of conducting electrons which cannot be reduced below $10^{17}$ $cm^{-3}$ by removing impurities. Doping with Sr remarkably reduces electrical conductivity of this material, that opens new possible applications, however, for obtaining highly resistive state (R~$10^9$ Ω) a high level of doping x~10% is necessary [30]. Characteristics of polycrystalline samples with lower level of SrO doping is absent in literature, and the reason of low efficiency of such doping in the frame of solid state technology was not discussed earlier.

Since the conductivity of materials with different level of doping depends differently on content of real air atmosphere (oxygen, hydrogen, humidity, vapours of ethanol, acetone etc.) a problem of comparative characterization of the samples arises. After preparation and cooling to room temperature, materials are in relatively high resistive state. The resistance of samples with highest resistivity exhibit remarkable changes with time when exposed to humid atmosphere. As was mentioned in previous section, the resistance drops sharply after mounting the electric contacts with conductive silver paint containing acetone. To restore the highly resistive state the samples were put after drying for a few minutes to vacuum and then annealed for 3 hours in oxygen atmosphere at temperature $300^0$C. Variation of the maximum sample resistance $R_{max}$ with SrO content is shown exemplarily by the curve 1 in Fig. 1.

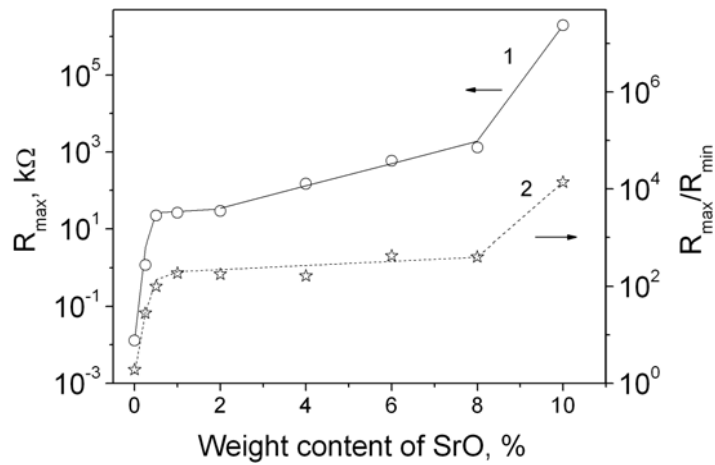

Fig. 1. Resistance of $In_2O_3$:$(SrO)_x$ samples in the highly resistive state $R_{max}$ (open circles and solid line) and its ratio to the resistance in the low resistive state $R_{min}$ (open stars an dashed line) depending on SrO content.

The second curve in Fig. 1 indicated by stars characterizes the range of the resistivity change between the high and the low resistive states by the ratio $R_{max}/R_{min}$. In this case, the low resistive state of samples was reached by oxygen content reduced in low vacuum ($10^{-2}$ bar) at temperature $300^0$C for 1 hour. The measurements were performed after 1 hour keeping the samples at room temperature following cooling down. It is apparent that reaching the low resistive state in a relatively short time is provided by changes in a near surface layer. Nevertheless, these data give an idea of sensitivity of material parameters to the oxygen content when the Sr content varies.

It is worth noting that both dependences shown in Fig. 1 exhibit special features in the SrO content regions x=0-1% and x=8-10%. With respect to pure indium oxide the resistance $R_{max}$ sharply increases by 3 orders of the magnitude below x=0.5%. $R_{max}/R_{min}$ increases in the region

x=0-1% by two orders of the magnitude. In the intermediate region, the resistance relatively slow increases further by two orders of the magnitude following roughly $R\sim exp(0.67x)$ during a substantial change of SrO content ($x$=1-8%). The corresponding change in the ratio $R_{max}/R_{min}$ is less than by a factor of 2. The next region of substantial changes is between $x$=8% and $x$=10% where the resistance increases by three orders of the magnitude while the ratio $R_{max}/R_{min}$ by approximately 1.5 orders of the magnitude. Thus the maximum $R_{max}/R_{min}$ variation of the order of $10^4$ presenting the highest interest for sensor applications is reached at maximum doping of $x$=10%. Further increase of the resistance makes no sense because the resistance $R_{max}$ values over $10^9\,\Omega$ can hardly be measured by standard technical means.

One of the reasons for high doping level of indium oxide with SrO necessary for reaching the highly resistive state has been established by means of X-ray investigations. It appeared that, additionally to the main $In_2O_3$ phase, another phase arises above $x$=2% which was identified as $SrIn_2O_4$ with a spinel structure (see Fig. 2). Structural properties of this phase have been investigated earlier [31]. No other phases containing Sr were established in the studied compounds.

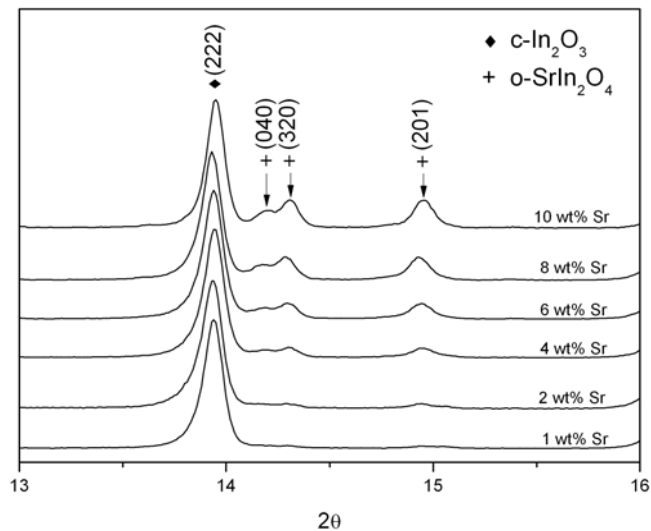

Fig. 2. XRD data for a compound resulting during synthesis of the strontium doped indium oxide. Intensity in all graphs is normalized to that of the largest peak from indium oxide at 2θ=13.98°. Reflection indicated with a symbol ♦ is from cubic $In_2O_3$ (space group Ia-3), reflections indicated by a symbol + are from orthorhombic $SrIn_2O_4$ (space group Pnam).

Useful information about the process of additional phase formation during solid state synthesis of the material was obtained by SEM investigations. SEM image in Fig. 3(a) shows the fragment of the sample surface with the boundary between the two phases. The main condition for the second phase growth during TTP is high initial concentration of $SrCO_3$ at a local position. Analysis of the element content by SEM service shows that relation of atoms concentration ($N_{Sr}/N_{In}$) in the lower region is close to 0.5 while in the upper one it is close to 0.02. It is thus clear that at a large local Sr content the second phase grows as a separate entity. $SrIn_2O_4$ phase is stable and, during the formation of its large grains, the material of numerous smaller neighbouring $In_2O_3$ grains is used. As a result of phase separation, Sr content is distributed very nonuniformly over the sample. This is one of the reasons of macroscopic inhomogeneity of electrical conductivity in the sample, because relatively high concentration of Sr in separate regions causes strong decrease of conductivity. To clarify electric properties of the additional phase formed during the solid state synthesis an additional sample was prepared with atomic content corresponding to the formula $SrIn_2O_4$. It was established that this sample was virtually dielectric with resistance $R>10^{12}\,\Omega$. It remained dielectric also after the thermal treatment at 300° C in vacuum.

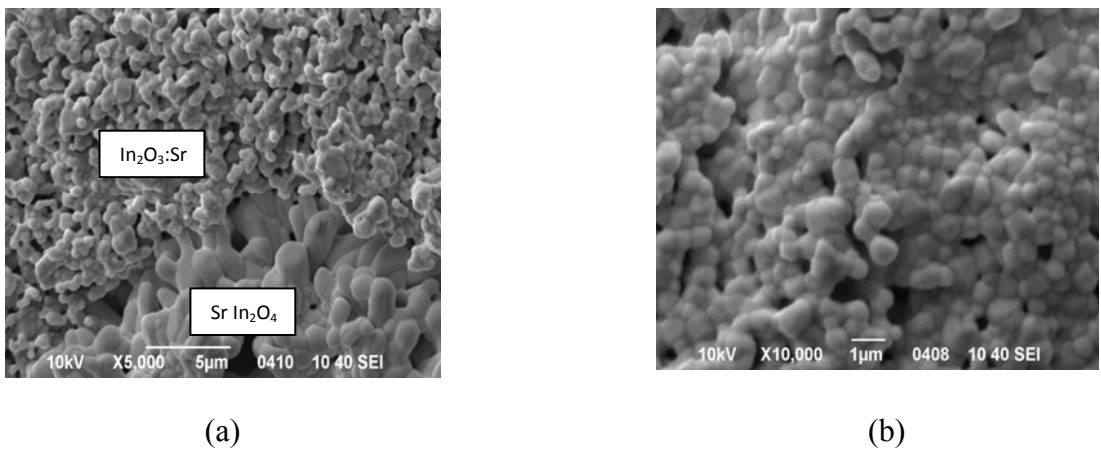

(a) (b)

Fig. 3. SEM image of the surface: (a) - of the compound $In_2O_3$:$(SrO)_x$ with $x=6\%$, displaying the formed spinel structure $SrIn_2O_4$ (lower part of the image) and the indium oxide grains containing much less Sr indicated here as $In_2O_3$:Sr (upper part of the image), and (b) - of the compound $In_2O_3$:$(SrO)_x$ with $x=10\%$.

Thus, the spatial picture of the sample conductance includes the microscopic dielectric regions. When the $In_2O_3$:$(SrO)_x$ material is far from percolation transition, direct influence of dielectric islands on the macroscopic conductance must be small. In the case of relatively small level of doping the second phase accumulate Sr atoms which do not distribute in the sample. As a result some part of Sr is electrically inactive for variation of macroscopic electrical properties. As was mentioned in introduction, the similar reason of low electric activity of Sn atoms was revealed in ITO films with doping level 6%. Electric inactivity of greatest part (2/3) of Sn was explained by their segregation at the surface. It can be supposed that at low doping the macroscopic conductivity of $In_2O_3$:$(SrO)_x$ is mostly affected by point defects, for example, Sr substitutions of In. At higher doping level the cells of the second phase, $SrIn_2O_4$, start to nucleate. A mechanism of incorporation of these cells into the indium oxide surrounding was considered in Ref. [31]. The strongest effect on the electric properties may result if the cells of the second phase are located at the grain boundaries of indium oxide and form potential barriers for electrical current. The presence of such barriers was established in Ref. [30].

## 4. Oxidation – reduction of the thick film samples in gas flow atmosphere

To test sensitiveness of the material to different atmospheres in view of possible sensor applications the electrical measurements were performed on thick film samples. The measurement procedure on thick film specimen with doping level of $x=10\%$ consisted of three stages: sample oxidation in pure oxygen atmosphere for 1 hour, the following reduction of oxygen content in pure nitrogen atmosphere for 2 hours, and the consequent long-lasting oxidation in pure oxygen atmosphere.

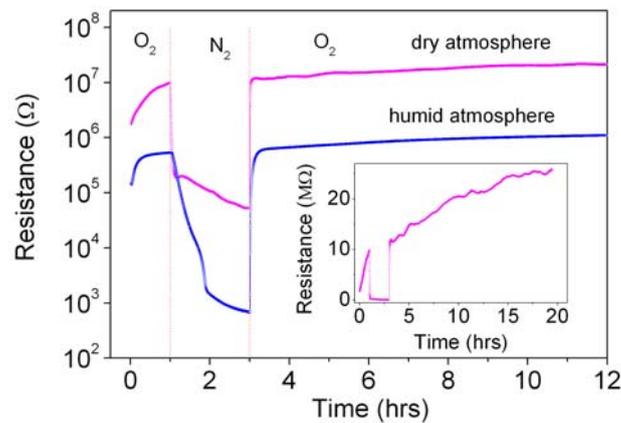

Fig. 4. Temporal variation of the thick film resistance at constant temperature $300^0C$: in pure oxygen atmosphere (first hour), in pure nitrogen atmosphere (next two hours) and in oxygen atmosphere (rest time). The upper and the lower curves present results in a dry and a humid atmosphere, respectively. The insert shows the data on the linear resistance scale for a longer duration.

As it is well seen from Fig. 4 (upper curve) the sample oxidation corresponds to an increase of resistance, and inversely, thermal treatment of the sample in nitrogen atmosphere causes a decrease of resistance. This is in accord with the concept of oxygen vacancies as shallow donor centers. The observed resistance behavior reveals, however, further subtle features. In Fig. 4, an existence of relatively fast and slow processes of resistance variation can be resolved implying different mechanisms involved. The speed of relative variation of sample resistance displayed in Fig. 5(a) demonstrates a big difference between time scales of these processes. Previously these time scales could be related to the processes at the surface and in the bulk of the grains [30].

The main process at the surface is chemisorption of oxygen, which creates surface acceptor states and decreases conductivity in a near surface layer. Density of acceptors states depends on partial pressure of oxygen and their variation in ambient atmosphere causes relatively fast reaction of the sample conductivity [32-34]. High sensitivity of thick film conductance due to the near surface layer is explained by relatively weak links between grains.

Slow process is supposed to be caused by diffusion of atomic oxygen in the grain interior. After changing the atmosphere content, the high-speed process produces a resistance variation of about two orders of the magnitude. The limit of high-speed resistance variation depends on the starting value of $R$. Slower variation of resistance at a larger time scale is clearly seen in the insert of Fig. 4.

The reaction of conductivity to oxygen content of atmosphere is a significant characteristic of materials for estimation their gas sensing ability [34]. The mechanisms of the observed conductivity variations might be related to various electrically active impurities which can be present in the material. First and foremost species to be considered are the water and hydrogen content. We can conclude in this respect that, according to results of Refs. [35,36], the temperature of 300°C is most appropriate for investigation of the oxygen content influence on the variation of the conductivity. Indeed, the water leaves indium oxide at lower temperature of 100-250$^0$C, while hydrogen get unbound at temperature of 320-400$^0$C. Thus we can suppose that bulk content of water in our thick films is minimal for the case of dry atmosphere.

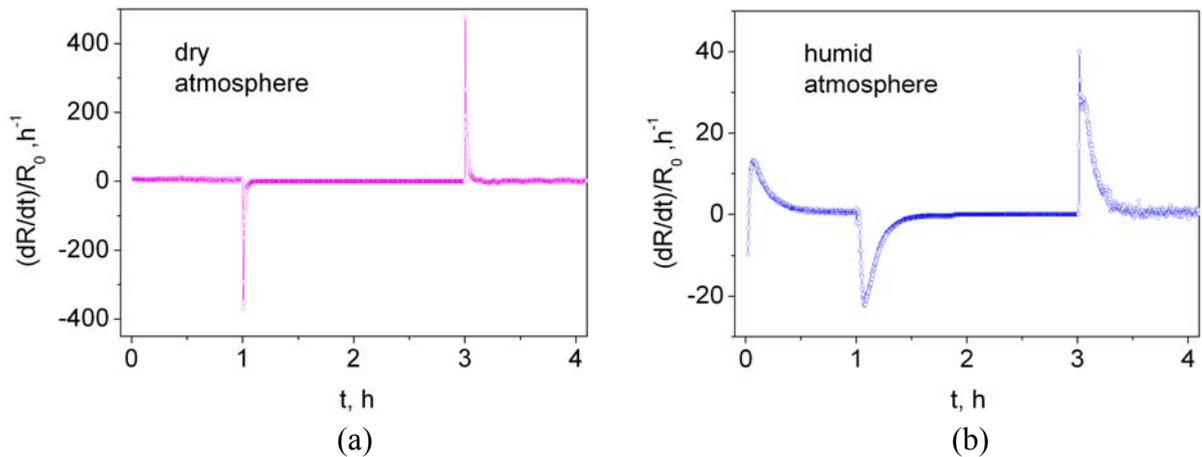

(a) (b)

Fig. 5. Temporal dependence of the speed of relative variation of sample resistance: (a) in a dry and (b) in a humid atmosphere.

The lower curve in Fig. 4 corresponds to the same conditions of measurements as for the upper curve but in a humid atmosphere. The difference of resistance between the two curves in dry and humid atmospheres is about one order of the magnitude for oxygen gas content and two orders for nitrogen. The influence of absorbed oxygen and humidity on the sample resistance is opposite, but the humid atmosphere provides one order greater variation of resistance in comparison with the dry one. It can be suggested that adsorbed water creates donor states. Except for this, as it is seen from Fig. 5(b), the presence of humidity decreases the speed and significantly increases the relaxation time of the absorption process. This, probably, reflects a competition between different molecules during absorption. On the other hand, the chemical dissociation of water molecules at the grain surface, as well as dissociation of oxygen and hydrogen molecules must be included to consideration of the absorption mechanism [33,34]. Additionally, hydrogen as shallow donor or passivation center for oxygen vacancy must be mentioned [37]. This, however, does not give a simple and complete explanation of action of humid atmosphere on the film resistance.

For comparison of sensitivity of thick film samples on the basis of $In_2O_3:(SrO)_x$ ($x = 10\%$) and undoped indium oxide the data of Ref. [34] can be used, where authors investigated the variation of conductance in dependence of oxygen content in Ar-$O_2$ gas atmosphere. The film sample was prepared by a technology similar to ours but using a special nano-size (13 nm) powder of pure $In_2O_3$ to increase sensitivity. According to these data, consecutive increase of oxygen content from zero up to 20% caused variation of film conductance by 8.6 times at film temperature $400^0C$. Extrapolation of this dependence allows estimation of the full variation of conductance by factor $\sigma_{max}/\sigma_{min}=17\div20$, when oxygen content of gas atmosphere changes from zero up to 100%. That is smaller than corresponding factors for fast resistance variation of Sr doped indium oxide in the actual work (see upper curve in Fig. 4). By the transition from pure nitrogen to oxygen atmosphere the fast resistance variation increases up to a factor of 220.

It must be noted that usually work temperature for gas sensing applications is in the range of $200\div500^0C$. Anorganic humidity meters are often used without heating of a sensor cell. The authors of Ref. [18] presented investigations of sensors on the basis of mixing compounds $In_2O_3$ and SiO. Sensitivity depended on temperature and was not very high. Particularly, at temperature $60^0C$ the changes of relative humidity from 35% to 90% caused the variation of sensor resistance of ~50-60%. High sensitivity of $In_2O_3:(SrO)_x$ ($x=10\%$) sample resistance to humidity is explained by not only specific properties of this material but also by the choice of working temperature of $300^0C$. Nevertheless, the questions concerning the contributions of bulk and surface conductivity mechanisms remain open.

## 5. Surface conductance and switching effect in polycrystalline $In_2O_3:(SrO)_x$

To obtain information on the low resistive state of the highly doped $In_2O_3:(SrO)_x$ with $x=10\%$ a procedure of short time (for 1h) thermal treatment in vacuum at $T=300^0C$ was previously used (see comments on Fig. 1 in Section 3). It was established that high conductivity is mostly provided by the oxygen reduction in the near-surface layer of the sample. In fact, this reveals an important material's ability of fast and considerable oxygen exchange with an ambient atmosphere. This property seems to be related to a specific friable structure of the crystallite surface.

Fig. 6 displays time development of the resistance of a polycrystalline sample of $In_2O_3:(SrO)_x$ ($x=10\%$) during its heating in low vacuum of $10^{-2}$ bar. The resistance starts to decrease fast after reaching temperature 50-70$^0C$ so that its magnitude decreases by 4 orders of the magnitude during one hour. During the subsequent cooling down to room temperature (not shown) its resistance grows by several times.

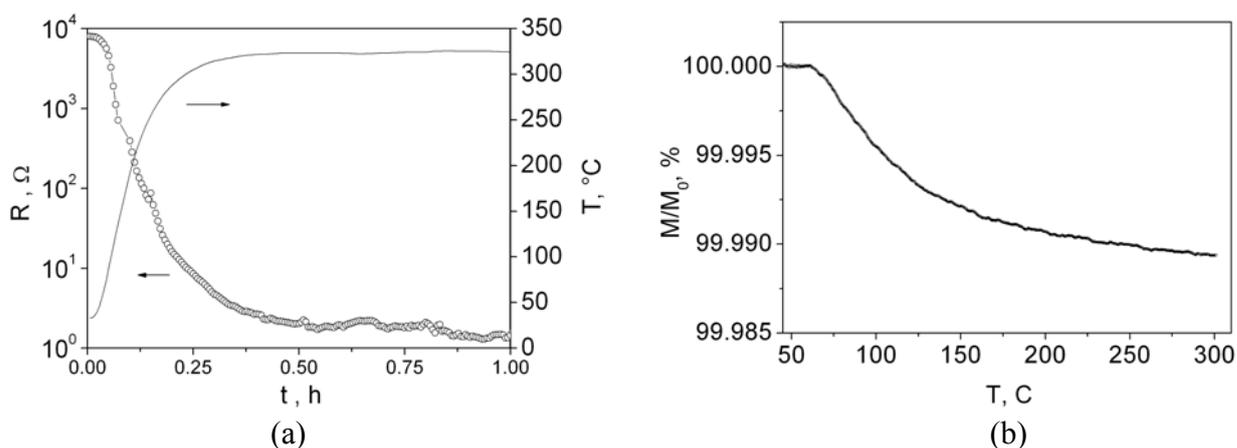

Рис. 6. (a) – Temporary development of the resistance of a $In_2O_3$:$(SrO)_x$ sample with (x=10%) during the increase of temperature in low vacuum ($10^{-2}$ bar). (b) - Variation of the sample mass during the heating in argon atmosphere.

To verify the oxygen variation quantitatively an additional gravimetric measurement was performed. Preliminary the polycrystalline sample was annealed at $300^0C$ for 12 hours in argon atmosphere and, after that, for 12 hours in oxygen atmosphere to provide strong decrease of humidity content and satiation of the sample with oxygen. The result of the sample mass measurement during the first half an hour in the regime of linear temporal increase of temperature up to $300^0C$ in the argon atmosphere is presented in Fig. 6(b). As it is well seen, the mass decrease first appears at temperature more than $60^0C$. Relatively significant and fast decrease of mass ($\sim 10^{-4}$ $\Delta m/m_0$) occurs in the temporal interval ~ 10-15 minutes during the increase of the sample temperature up to $150-200^0C$. This process can be related to oxygen extraction from the sample surface promoted by the polycrystalline structure and porosity. Actually, the density of oxygen atom loss estimated from the weight loss is approximately $2.65 \times 10^{19}$ $cm^{-3}$. Taking into account the grain diameter of 300 nm the total grain surface in the sample with a volume of 1 $cm^3$ amounts to about $10^5$ $cm^2$. Since the volume corresponding to the chemical unit of $In_2O_3$ is about 1 $nm^3$, the density of oxygen atoms in the surface layers of all grains to a depth of 1 nm can be estimated as $3 \times 10^{19} cm^{-3}$. Therefore, the quantity of oxygen atoms in the sample rapidly lost during the heating procedure is close to the full oxygen content in the surface layer (with account of granular structure) of the depth of 1 nm. If one assumes the existence at the grain boundaries of a friable structure consisted of disordered and incomplete crystal cells this may explain a significant and fast oxygen exchange with environment. Note that during the subsequent thermal treatment of the sample for long time (30 hours) the sample mass loss was estimated by the value of only 4 times greater.

To be able to judge on properties of very thin layer of the polycrystalline $In_2O_3(SrO)_x$ an insight in processes at the microscale is required. To additionally characterize the surface processes and to establish properties of the hypothetical friable structure an investigation of the diffusion mobility of electrically active defects in the near surface layer of the sample has been performed. After extraction of oxygen in vacuum the sample was connected to the electrical circuit by two contacts of different areas (see Fig. 7(a)). A bottom plate contact with a large area was characterized by relatively small resistance (~100 Ω). The top contact was realized with the help of a pressed metal tip. It was characterized by relatively high resistance (~1MΩ). Thus, the measured resistance was dominated by the properties of the small-area contact. The electrical circuit (Fig. 7(a)) supplied the sample in the regime of constant voltage and provided the possibility of step-like switching of voltage polarity.

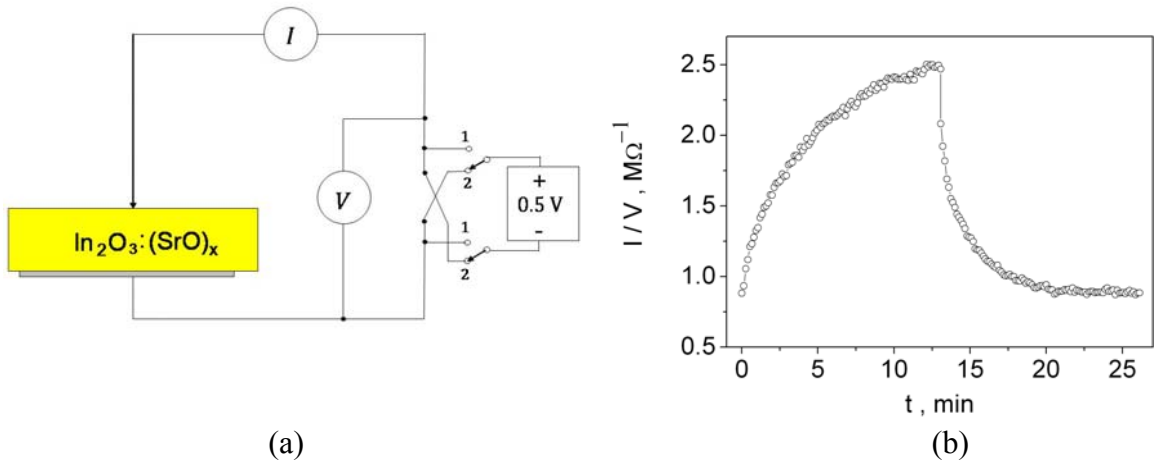

(a) (b)

Fig. 7. (a) Electrical scheme for controlling the switching effect. (b) Fragment of the temporal dependence of the conductance of a small area contact (tungsten pin) after multiple periodical step-like switching of voltage polarity.

It was revealed that each switching of voltage polarity causes variation of the small area contact conductance with time on the typical time scale of 10 minutes (see Fig. 7(b)). To understand the nature of this phenomenon the main conditions of its observation were investigated. The phenomenon is not much sensitive to a kind of metal used in electrodes. Similar results were observed for different metal needles of stainless steel, tungsten and gold. The results of the experiment with the gold pin exclude relation of the switching effect nature to electrochemical processes at the metal tip. Hence, the switching effect is caused by the process in the sample vicinity near the electrical contact. Furthermore, the effect was not observed when a small contact of about 0.1 mm$^2$ area was prepared by means of magnetron sputtering of Pt target. The reason for this can be related to the substantial increase of the area compared with the tip contact area and the consequent decrease of the small contact resistance. In principle, this effect must be compensated in the case of two identical contacts, because the temporal variation of resistance is then opposite for the two contacts at different voltage polarity. This effect was not observed either in the case of pure $In_2O_3$ for temporal scale of several minutes. Undoped indium oxide was characterized by low resistive state, and the sample was not annealed in vacuum before the measurement.

It should be noted, that the switching effect in $In_2O_3$:$(SrO)_x$ with x=10% during serial repeating of measurements is described by well reproducible curves σ(t) (see Fig. 7(b)) only in the case of not too large duration of measurement cycles (several minutes or less). If the duration between polarity switching is half an hour or more, each next curve becomes slightly different. One of the features of this difference is a monotonic shift of the average value of resistance. This fact attests to complexity of resistance relaxation mechanisms revealed on the different time scales.

Taking into account that the switching effect is observed in the low resistive state of In$_2$O$_3$:(SrO)$_x$ after strong extraction of oxygen in vacuum at 300$^0$C, a scenario of the relevant physical process can be suggested. Since the surface of grains located close to the sample surface is characterized by larger oxygen deficit than in the bulk, the diffusion mobility of oxygen anions is expected to be significantly greater than in the stoichiometric bulk material. The next reason for the mobility increase may be a poorly packed friable structure of the grain surface layer. As a result, vacancies move under the action of the electrical field and change the local density of free carriers in the vicinity of the small area contact where the electric field is large.

For a semi-infinite sample with a homogeneous conductivity the resistance of the small area contact of a semi-spherical form of the tip can be expressed as $R=\rho/2\pi r_0$ where $\rho$ is resistivity and $r_0$ is the radius of the semi-spherical tip. Correction to this formula due to a realistic two-electrode geometry is as small as $r_0/w$ where $w$ is the lateral size of the bottom electrode. Hence, the resistance of the small area contact is affected mainly by the contact vicinity of the order of the radius $r_0$. If the specific resistances of the sample material is about 50 Ω·cm and resistance of tip contacts is about R~ 1 M Ω, the value of $r_0$ ~80 nm can be estimated in our experiment. This gives a size scale of the layer which can be judged on from this experiment.

Calculation of the resistance variation can be performed by taking into account several physical mechanisms: a direct electric field force, an effect of electron wind and oxygen vacancies diffusion flow. Because the electron wind interacts with oxygen anions and do not affect directly their vacancies, the electric field force and the electron wind move oxygen vacancies in the same direction. Diffusion mechanism prevents large increase of vacancies density. An appropriate quantitative model is formulated and solved in the next section.

**6. Nonlinear drift-diffusion model of conductivity**

For description of the electric conduction process in oxygen deficient systems one has to account for slow evolution of the local system properties due to migration of oxygen vacancies subject to an applied electric field. We assume that, driven by the voltage applied to the (plate or pin) electrode, oxygen vacancies redistribute in space according to the continuity equation for the vacancy concentration C(**r**,t) supplemented by the expression for the drift-diffusion current density of vacancies **J**(**r**,t) :

$$\partial C / \partial t = -\nabla \mathbf{J} \qquad (1)$$

$$\mathbf{J} = (qD/k_B T)C\mathbf{E} - D\nabla C \qquad (2)$$

where $D$ is the diffusion coefficient of oxygen vacancies with effective charge $q$, which characterizes an action of both the electric field **E** and the electron wind. The Einstein relation between the diffusivity and mobility of vacancies was also implied in Eq. (2).

Electrons, as much more mobile species than oxygen vacancies, are supposed to completely and instantly screen the positively charged vacancies so that the local electroneutrality prevails and the relation for the local electron concentration $n \sim C$ holds.

Accordingly, the local neutrality condition requires that $\nabla(\mathbf{j} + q\mathbf{J}) = 0$ where $\mathbf{j}$ is the electron current. However, since the contribution of electrons to the electric current is by orders of the magnitude larger than that of oxygen vacancies, the latter can be neglected. This results in the continuity equation for the electrons alone supplemented by the Ohmic law with the local conductivity value dependent on the local concentration of oxygen vacancies

$$\nabla \mathbf{j} = 0 \quad \text{with} \quad \mathbf{j} = \sigma(C)\mathbf{E}. \tag{3}$$

The conductivity is expected to increase with the local density of donors as $n \sim C$ so that we simply adopt $\sigma(C) \sim n \sim C$. Here we have accounted for the fact that, after doping with Sr, the density of electrons in $In_2O_3$ was reduced by about 6 orders of the magnitude which can be inferred from the correspondent increase in resistance. Keeping in mind the concentration of free electrons of $10^{17}$ cm$^{-3}$ in undoped $In_2O_3$ [4] and the above evaluated density of oxygen vacancies of about $10^{19}$ cm$^{-3}$ in oxygen deficient $In_2O_3:(SrO)_x$, we assume that, in the latter compound, the major contribution in the number of electrons is due to oxygen vacancies as donors.

The system of equations (1-3) applies to the one-dimensional geometry of a sample between two plate contacts as well as to the asymmetric case with a needle contact at one side where much stronger effect on resistance due to migration of oxygen vacancies is expected. Leaving the details of numerical treatment for a separate publication, the results of the numerical solution of the above system of equations for the asymmetrical case are presented in Fig. 8.

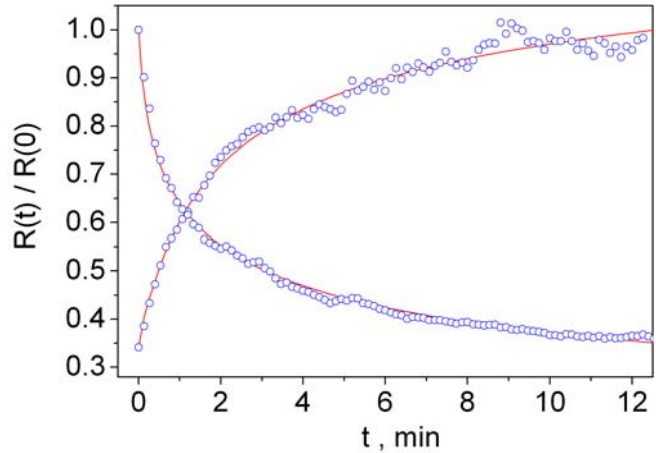

Fig. 8. Temporal dependencies of the relative variation of the resistance of a small area contact for the case of negative (descending curve) and positive (ascending curve) polarity of voltage on the metal tip. Open circles present the experiment with a tungsten tip, solid lines - the theory.

The best fitting of experimental data $R(t)/R(0)$ with theoretical curves was obtained for two parameters concerning oxygen vacancies, their effective charge and diffusivity: $q \approx 0.3e$ and $D \approx 3 \times 10^{-13}$ cm$^2$/s (with $e$ the elementary charge). It should be noted, that the estimated value of $D$ is by one order of the magnitude smaller, than the corresponding value for amorphous films estimated in Ref. [38]. In principle, this is in agreement with our expectation taking into account the concept of the friable surface structure.

## 7. Discussion of results

The main feature of the strontium doped indium oxide $In_2O_3$:$(SrO)_x$, prepared by a solid state technology and studied in the current work, is a monotonic but non-uniform increase of the electrical resistance by 8 orders of the magnitude during the doping increase from $x=0$ to 10% accompanied by the change in its structure and physical properties. The fastest increase in the resistance is observed in the regions $x=0-0.5\%$ and $x=8-10\%$ (Fig. 1). Starting with $x=2\%$ a formation of an additional $SrIn_2O_4$ phase with a spinel structure was identified by means of XRD (Fig. 2) and SEM (Fig. 3) analysis.

Taking into account relatively low strontium content in bulk samples it is not expected that the presence of microscopic local inclusions of the second, dielectric phase (Fig. 3(a)) may be a major reason for the reduction of the electric conductivity in the doped material (Fig. 1). Hence, this effect should be explained by the diffusive redistribution of a smaller Sr amount within the main, conductive phase.

Based on the fact that, during the thermal treatment at $1200^0C$, the grains of the second phase succeed to reach the size remarkably larger than that of $In_2O_3$ grains (Fig. 3(a)), a conclusion on the sufficiently high diffusion mobility of cations follows which concerns also the volume of grains. Since the surface diffusivity of atoms substantially exceeds the volume one, it is expected to provide a macroscopically uniform distribution of strontium over the sample while the volume diffusivity is responsible for its distribution within grains. Homogenous distribution is, however, prevented by the formation and stability of the $SrIn_2O_4$ phase. From the elemental analysis (EDX) of indium oxide areas of the $In_2O_3$:$(SrO)_x$ samples with nominal (precursor) SrO content of $x=6\%$ (Fig. 5(a)) it was established that the main indium oxide phase contains no more than 2% of Sr.

In Fig. 3(b), a surface fragment of the sample with $x=10\%$ is displayed. No large inclusions of the second phase are observed in this area. The material is porous but the grains with a typical size of 0.3-0.5μm are well merged and as a rule smoothly contacted. The following possible mechanisms of the conductivity reduction in the main phase are conceivable. Firstly, the acceptor role of strontium as an element of the second group is straightforward when substituting indium. Secondly, a formation of regions with different strontium contents is possible according to scenario of Ref. [31]. A related notion of the "inter-crystalline network" has been used by the authors of Ref. [39] to explain properties of polycrystalline materials with $x=10\%$. The third option is a formation on the surface of indium oxide grains of a thin tunnel-transparent or low conducting layer with a possible participation of the second phase.

Coming back to the data in Fig. 1 it can be assumed that, at low level of doping ($x=0-0.5\%$), strontium reduces conductivity mostly according to the first and the second mechanisms. In the intermediate phase ($x=1-8\%$), the increase of the resistance is impeded because of the formation of local regions of the second, dielectric phase. Here, presumably, the second phase absorbs significant part of the increasing Sr content. The presence of local dielectric inclusions hardly affects the conductivity of the conducting matrix because the system is still far away from the percolation threshold. On the other hand, the structure of the material is microscopically non-uniform. In the granular medium there are inter-granular contacts which strongly influence the macroscopic conductivity of the material and are prone to diffusion penetration of strontium. It may be supposed that the sharp increase of the resistance in the region $x=8-10\%$ is related to a lost of percolation through the conducting network of the microscopically inhomogeneous structure.

Being on the brink of the percolation transition the system becomes highly sensitive to other factors influencing surface properties, for example, gas adsorbtion from the ambient atmosphere. This sensitivity is growing with the increase of the electrical resistance of the material. The study of the fine grained powder samples with $x=10\%$ in a gas flow atmosphere exhibits a high and fast susceptibility of the resistance to the oxygen content and humidity (Fig. 4). Temporal characteristics of the resistance variation displayed in Figs. 4 and 5 demonstrate fast adsorbtion processes at the time scale of seconds and minutes as well as slow diffusion processes at the scale of dozens of hours. The presence of humidity retards oxygen exchange with the surrounding that can be explained by chemical reactions involving water at the grain surfaces.

A remarkable property of the highly resistive $In_2O_3:(SrO)_x$ compounds with $x=10\%$ is an anomalously fast and quantitatively large oxygen exchange with the ambient atmosphere (Fig. 6). This fact seems to indicate a specific friable structure of the crystallite surface. The formation of this surface structure is related to mechanism of strontium integration into indium oxide crystallites [31]. The depth of friable structure can be increased by means of additional thermal treatment procedure in vacuum that will be published elsewhere. Providing much higher oxygen diffusivity such a structure promotes a relatively fast transition of the sample from high to low resistive state at the expense of high electrically conducting near surface layer of the crystallites.

Microscopic study of electrodiffusion processes at the surface of oxygen deficient $In_2O_3:(SrO)_x$ sample with $x=10\%$ (Fig. 7) allowed estimation of the diffusion coefficient of oxygen vacancies using an original nonlinear drift-diffusion model. The estimated value of $D=3\times10^{-13}$ cm$^2$/s at room temperature is one order smaller than the value known for amorphous indium oxide films [38]. This fact can be considered as additional evidence of existence of the friable surface layer of crystallites.

## 8. Conclusions.

Sr doped indium oxide compound $In_2O_3:(SrO)_x$ with doping $x$ up to 10% fabricated by solid state processing was shown to possess exceptional properties making it interesting for sensing applications. Increasing Sr content from $x=0$ to $x=10\%$ monotonically elevates resistivity of the material by eight orders of the magnitude. The resistivity raise is not uniform. The first increase occurs within the range $0<x<0.5\%$ and is related to the acceptor role of Sr substituting In. The second increase occurs in the range $8\%<x<10\%$ and exhibits features of percolation transition to a dielectric phase. The weak resistance increase in the intermediate range $1\%<x<8\%$ is explained by the formation of the second, dielectric phase $SrIn_2O_4$ with a spinel structure, observed by both XRD and SEM, which mostly absorbs additional Sr. The higher is the doping level $x$ the stronger is sensitiveness of the material to the contents of the ambient atmosphere.

Heat treatment of different compounds at 300°C in vacuum, argon, nitrogen and humid atmosphere shows that maximum variation of the resistance changes from the factor of two for $x$=0.25% to four orders of the magnitude for $x$=10%. Large and fast variation of the resistance in response to the changing atmosphere content suggests sensing applications of the latter compound.

Strong variation of the resistance of highly doped (and highly resistive) samples is explained by oxygen exchange with the ambient atmosphere with the maximum resistance corresponding to the highly oxided and minimum resistance corresponding to the reduced state. High oxygen mobility in $In_2O_3$:$(SrO)_x$ is confirmed by both thermogravimetry in the flux-flow atmosphere and by local measurements of the resistivity using small area pin contact. The latter measurement exhibits switching effect of resistivity at the time scale of several minutes and allows determination of the local diffusivity of oxygen vacancies in the area near the contact tip at room temperature which amounts to $D=3\times10^{-13}$ cm$^2$/s. This relatively high value characterizes the surface region of the indium oxide microcrystallites with a specific friable structure due to both the strontium integration into crystallites and oxygen exchange with the ambient atmosphere.

**Acknowledgements**

We recognize providing a facility for point contact measurements with different metal pins and useful discussions with Prof. A. Klein. This work was partly supported by the Deutsche Forschungsgemeinschaft through the Sonderforschungsbereich 595 "Electrical Fatigue in Functional Materials".